# A comparative study of the structural, elastic, thermophysical, and optoelectronic properties of $CaZn_2X_2$ (X = N, P, As) semiconductors via *ab-initio* approach

Md. Sajidul Islam, Razu Ahmed, Md. Mahamudujjaman, R.S. Islam, S.H. Naqib*
Department of physics, University of Rajshahi, Rajshahi 6205, Bangladesh
*Corresponding author email: salehnaqib@yahoo.com

**Abstract**
We present a detailed density functional theory (DFT) based calculations of the structural, elastic, lattice dynamical, thermophysical, and optoelectronic properties ternary semiconductors $CaZn_2X_2$ (X = N, P, As) in this paper. The obtained lattice parameters are in excellent agreement with the experimental values and other theoretical findings. The elastic constants are calculated. These elastic constants satisfy the mechanical stability criteria. Moreover, many thermophysical parameters of these compounds are estimated, including the Debye temperature, average sound velocity, melting temperature, heat capacity, lattice thermal conductivity, etc. The comprehensive analysis of the elastic constants and moduli show that $CaZn_2X_2$ (X = N, P, As) compounds possess reasonably good machinability, relatively high Vickers hardness and relatively low Debye temperature. The phonon dispersion curves and phonon density of states are investigated for the first time for the compounds $CaZn_2P_2$ and $CaZn_2As_2$. It is observed from the phonon dispersion curves that the bulk $CaZn_2X_2$ (X = N, P, As) compounds are dynamically stable in the ground state. Electronic properties have been studied through the band structures and electronic energy density of states. HSE06 (hybrid) functional is used to estimate the band gaps accurately. The electronic band structures show that $CaZn_2N_2$ and $CaZn_2As_2$ possess direct band gaps while the compound $CaZn_2P_2$ show indirect band gap. It is observed that the band gap decreases by changing the anion X from N to As. The bonding characters of $CaZn_2X_2$ (X = N, P, As) compounds are investigated. Energy dependent optical parameters exhibit good correspondence with the electronic energy density of states features. We have thoroughly discussed the reflectivity, absorption coefficient, refractive index, dielectric function, optical conductivity and loss function of these semiconductors. The optical absorption, reflectivity spectra and the refractive index of $CaZn_2X_2$ (X = N, P, As) show that the compounds hold promise to be used in optoelectronic devices.

**Keywords:** Ternary semiconductors; Density functional theory; Elastic properties; Thermophysical properties; Optoelectronic properties

## 1. Introduction

In recent times the semiconductor industry drives the engine of condensed matter physics and materials science. Ternary semiconductors have piqued the interest of modern electronic industries due to their favorable electronic features and wide range of applications in optoelectronic and electronic devices [1–3]. The III-V group nitride, phosphide and arsenide are promising systems for light-emitting diodes (LEDs), laser diodes and solar cells.

Nowadays, LEDs and laser diodes are prepared from III-V nitride, phosphide and arsenide semiconductors [4]. Nitride semiconductors have been widely studied for a variety of applications. The main group metal nitride semiconductors with wurtzite crystal structure are used in various optoelectronic devices. These materials have become important over the last few decades due to the rapid increase of the demand for solid-state lighting, radio frequency transistors and high information density optical storage media [5,6]. The nitride semiconductors of the group III elements are commercially available for the production of high-efficiency solid-state devices that emit green and blue light [6]. Besides, these materials can be applied in flat-panel displays, high density optical data storage and high-resolution printing [6]. Moreover, the Zn-based nitride semiconductors would be promising candidates for the next generation functional optoelectronic semiconductors because they have similar desirable electronic properties to the well-known semiconductors GaN and InN [7]. Among these $CaZn_2N_2$ and its alloy system $Ca(Mg_{1-x}Zn_x)_2N_2$ (x = 0 to 1) have small electron and hole effective masses, and they show tunable direct band gaps which can be continuously varied from 3.2 to 1.8 eV [8,9]. These materials can also show band to band photoluminescence at room temperature in the ultraviolet-red region depending on the value of x. When x = 0.5, the specimen displays intense green emission with a peak at 2.45 eV [9]. From these features of $Ca(Mg_{1-x}Zn_x)_2N_2$, it can be said that the III-V group nitride, phosphide and arsenide semiconductors can be used in LEDs and in the absorption layers of solar cells. In addition, many nitride semiconductors are favorable for applications in power electronics [8]. The ternary phosphide compound $CaZn_2P_2$ and their alloys have good carrier mobilities [10]. The phosphide alloy $Sr(Zn_{0.5}Cd_{0.5})_2P_2$ has an ideal band gap of 1.38 eV and high absorption coefficient [10]. It contains Cd element that is less toxic and it is a promising material for optoelectronic applications [10].

In this study, the physical properties of three promising semiconducting compounds $CaZn_2X_2$ (X = N, P, As) [7,11] with space group $P\bar{3}m1$ belonging to the trigonal system, have been investigated. In this paper we cover the structural, elastic, thermophysical and optoelectronic properties of these three compounds. The results obtained have been analyzed by comparing with the available experimental data and other theoretical findings. It should be noted that many of technologically pertinent physical properties of $CaZn_2X_2$ (X = N, P, As) compounds remain comparatively unexplored compared to other systems such as oxides, chalcogenides and pnictides.

Recently, hetero-epitaxial $CaZn_2N_2$ thin films were synthesized experimentally by Tsuji et al. [4]. Furthermore, the energy band gaps of $CaZn_2N_2$ and other related compounds were studied theoretically as well as experimentally by Hinuma et al. [8]. From the absorption spectra plots it was found that $CaZn_2N_2$ has a band gap of 1.9 eV. Murtaza et al. [11] theoretically discussed the structural, electronic and optical properties of the five Zintl compounds $CaZn_2X_2$ (X = N, P, As, Sb, Bi). The systematic variation of anions X from N to Bi was reported therein. The thermodynamic properties of pure and doped $CaZn_2P_2$ compounds were characterized experimentally by Phonnambalam and Morelli [12]. From the calculated figure of merit and transport measurements they concluded that the $CaZn_2P_2$ and $CaMnZnP_2$ are high temperature thermoelectric materials. Hua et al. [13] theoretically predicted the electronic and magnetic nature of Mn and Na co-doped $CaZn_2As_2$ and observed that Mn doped $CaZn_2As_2$ shows antiferromagnetic behavior. There are few other studies on the Zintl phases compounds $CaZn_2X_2$ and also on the related systems [7,14].

Among the three compounds considered in this study, experimental studies have been done on the compounds $CaZn_2N_2$ and $CaZn_2P_2$ [4,12] but the compound $CaZn_2As_2$ has not been experimentally synthesized yet to the best of our knowledge. There are some theoretical studies on the elastic properties of the compound $CaZn_2N_2$ but no notable work has been reported yet on the elastic properties of the compounds $CaZn_2P_2$ and $CaZn_2As_2$. Moreover, the lattice dynamical properties of the compounds $CaZn_2P_2$ and $CaZn_2As_2$ have not been studied at all. There is also a lack of information on the various thermophysical properties of these compounds such as Debye temperature, lattice thermal conductivity, melting temperature etc. Besides, still there are notable shortages of theoretical understanding of the bonding and optical properties of these compounds.

From the above discussion, it is clear that many physical properties of these materials are waiting to be investigated. Therefore, it is of great theoretical interest to investigate these unexplored properties. Motivated by the lack of fundamental data, in this work our aim is to cover the structural, elastic, lattice dynamical, bonding, optoelectronic and thermophysical properties of $CaZn_2X_2$ (X = N, P, As) semiconductors in detail.

The rest of this paper is arranged as follows: we briefly describe our computational methodology in Section 2. In Section 3, we presented the obtained results and discussion of various properties of $CaZn_2X_2$ (X = N, P, As) semiconductors in detail. Finally, in Section 4 important findings of this work are summarized.

**2. Computational methodology**

The ground state physical properties of the $CaZn_2X_2$ (X = N, P, As) compounds were explored by means of density functional theory (DFT) [15] as implemented in the CASTEP code [16]. In DFT formalism, the ground state of a crystalline system is determined by solving the Kohn-Sham equation [17]. In this study, we choose the generalized gradient approximation within the Perdew-Burke-Ernzerhof (GGA-PBE) [18] scheme for structural optimization since it provides better results of ground state structural parameters for $CaZn_2X_2$ than the local density approximation (LDA) [19]. Vanderbilt-type ultra-soft pseudopotential [20] has been used to model the electron-ion interaction. Ultra-soft pseudopotential saves significant amount of computing time with a minimal loss of computational accuracy. The BFGS (Broyden-Fletcher-Goldferb-Shanno) minimization approach has been employed to find the ground state of the crystal atructure [21].

To perform the pseudo atomic computations, the valence electron configurations of Ca, Zn, N, P and As atoms have been taken as [$3s^23p^64s^2$], [$3d^{10}4s^2$], [$2s^22p^3$], [$3s^23p^3$] and [$4s^24p^3$], respectively. In this work, 7×7×4 k-points mesh is used based on the Monkhorst-Pack scheme [22] for sampling the Brillouin zone of $CaZn_2X_2$ compounds which results in a high degree of total energy convergence. A cut-off energy of 400 eV is applied for $CaZn_2N_2$ compound while for $CaZn_2P_2$ and $CaZn_2As_2$ a cut-off energy of 500 eV is used. The structure is relaxed up to a convergence threshold for energy of 5×10$^{-6}$ eV-atom$^{-1}$, maximum force of 0.01 eV Å$^{-1}$, maximum stress of 0.02 GPa, and a maximum atomic displacement of 5×10$^{-4}$ Å.

The independent elastic constants $C_{ij}$, bulk modulus $B$, shear modulus $G$, and other elastic properties are calculated for the optimized geometry. The Heyd–Scuseria–Ernzerhof (HSE06) hybrid functional [23] is used to investigate the electronic band structure, total and partial density of states (TDOS and PDOS, respectively) and bond population analysis of the CaZn$_2$X$_2$ compounds. Various thermophysical properties such as Debye temperature, melting temperature, thermal conductivity, heat capacity etc. are calculated from the obtained elastic parameters.

The optical characteristics of CaZn$_2$X$_2$ (X = N, P, As) compounds are explored in the ground state using the complex dielectric function, $\varepsilon(\omega) = \varepsilon_1(\omega) + i\varepsilon_2(\omega)$. The real part [$\varepsilon_1(\omega)$] of the dielectric function can be derived from the imaginary part $\varepsilon_2(\omega)$ by using the Kramers-Kronig relationships. The imaginary part, $\varepsilon_2(\omega)$, is determined from the electronic band structure by applying the CASTEP supported formula, which can be expressed as [24]:

$$\varepsilon_2(\omega) = \frac{2e^2\pi}{\Omega\varepsilon_0}\sum_{k,v,c}|\langle\psi_k^c|\hat{u}.\vec{r}|\psi_k^v\rangle|^2\,\delta(E_k^c - E_k^v - E) \qquad (1)$$

In this equation, $\Omega$ is the volume of the unit cell, $\omega$ is frequency of the incident electromagnetic wave (photon), $e$ is the electronic charge, and $\psi_k^c$ and $\psi_k^v$ are the conduction and valence band wave functions at a given wave-vector $k$, respectively. The delta function enforces energy and momentum conservation during the photon induced transition of the electron from the valence to the conduction band. Once the frequency/energy dependent dielectric function $\varepsilon(\omega)$ is known, it can be used to calculate the other optical properties such as the conductivity, refractive index, absorption coefficient, reflectivity, and energy loss function using the standard formalism.

## 3. Results and discussions

### 3.1. Structural properties

The geometry optimization results in the fully relaxed lattice parameters of the chosen compounds. The compounds CaZn$_2$X$_2$ (X = N, P, As) have a trigonal crystal structure with hexagonal symmetry. The space group associated with CaZn$_2$X$_2$ is P$\bar{3}$m1 (no.164) [11]. This structure contains one formula unit per unit cell. The conventional two and three dimensional unit cell of CaZn$_2$X$_2$ is depicted schematically in Figure 1. The X anions form an almost undistorted hexagonal close packed array; the Ca atoms cover one layer of the octahedrally coordinated locations, while the Zn atoms occupy all tetrahedrally coordinated positions in the next layer.

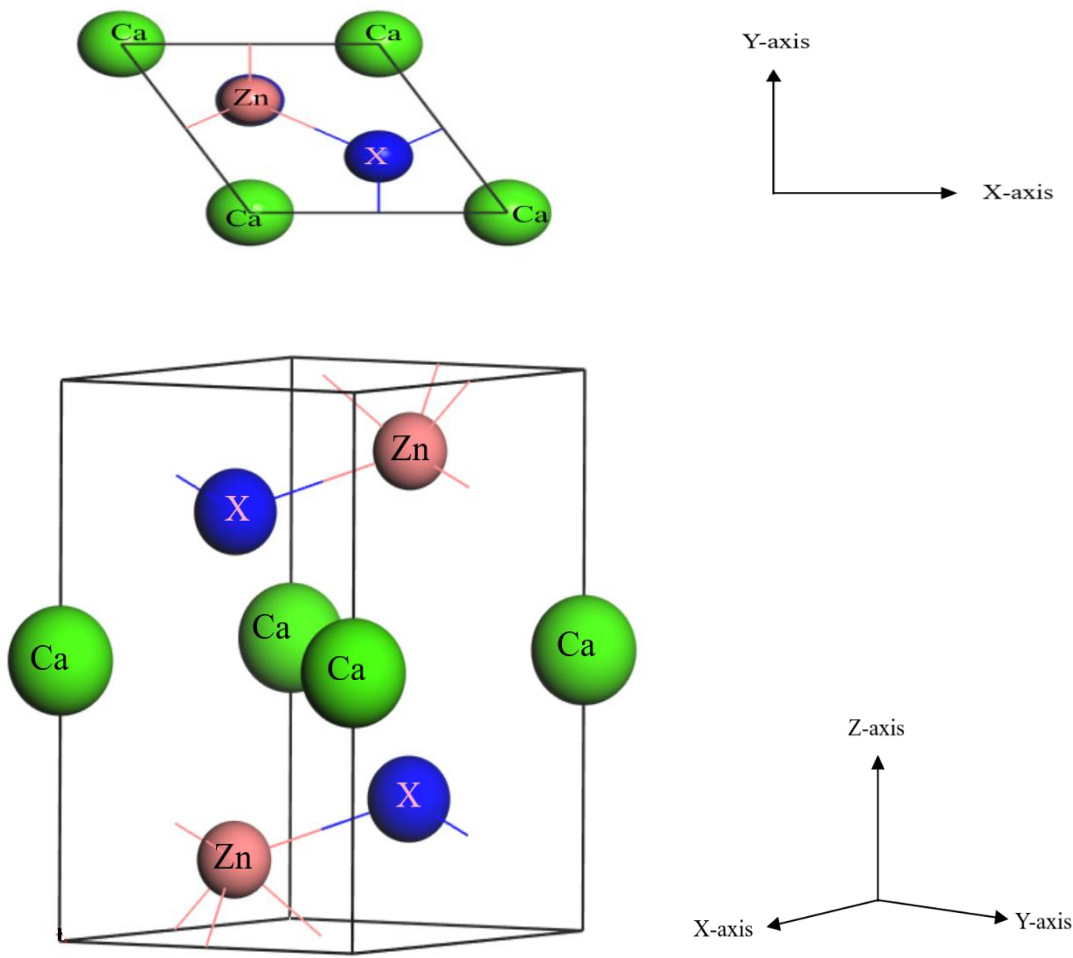

**Figure 1.** Two and three dimensional schematic view of the unit cell of $CaZn_2X_2$ (X = N, P, As) compounds.

The lattice parameters obtained from the geometry optimization along with available experimental and theoretical data are listed in Table 1 for comparison. The results obtained by the functional GGA(PBE) differ from the experimental values by less than 0.36%, 0.48% and 0.95% for *a*, *c* and $V_0$, respectively. These results indicate that they are in excellent agreement with the experimental data. These results are also in reasonable agreement with the theoretical findings. On the other hand, the functional LDA tends to underestimates the equilibrium structural parameters and the results deviate from the reported experimental data by greater than 2.40%, 2.44% and 7.06% for *a*, *c* and $V_0$, respectively. For this specific reason, we choose the functional GGA(PBE) to calculate the elastic/mechanical/thermophysical properties of the of $CaZn_2X_2$ compounds. Again, from Table 1 it is revealed that the lattice parameter increases as we change the anions X from N to As. This happens due to the replacement of small ionic size atoms with atoms having larger ionic size.

**Table 1.** Calculated lattice parameters ($a = b$ and $c$ in Å, equilibrium volume $V_0$ in Å$^3$) of the CaZn$_2$X$_2$ (X = N, P, As) compounds as compared to available experimental (Expt.) and theoretical (Theo.) values.

| Compound | Functionals | $a$ | $c$ | $c/a$ | $V_0$ | Ref. |
|---|---|---|---|---|---|---|
| CaZn$_2$N$_2$ | GGA(PBE) | 3.471 | 6.038 | 1.740 | 63.002 | This work |
| | LDA | 3.380 | 5.863 | 1.735 | 57.999 | This work |
| | - | 3.463 | 6.009 | 1.735 | 62.408 | [4] Expt. |
| | GGA(PBE) | 3.490 | 6.018 | 1.724 | 63.460 | [11] Theo. |
| CaZn$_2$P$_2$ | GGA(PBE) | 4.034 | 6.828 | 1.693 | 96.206 | This work |
| | LDA | 3.928 | 6.631 | 1.688 | 88.584 | This work |
| | - | 4.038 | 6.836 | 1.693 | 96.531 | [12] Expt. |
| | GGA(PBE) | 4.040 | 6.823 | 1.687 | 98.25 | [11] Theo. |
| CaZn$_2$As$_2$ | GGA(PBE) | 4.177 | 7.015 | 1.679 | 106.006 | This work |
| | LDA | 4.048 | 6.791 | 1.678 | 96.390 | This work |
| | - | 4.162 | 7.010 | 1.685 | 105.161 | [25] Theo. |
| | GGA(PBE) | 4.190 | 7.079 | 1.690 | 107.540 | [11] Theo. |

## 3.2. Elastic properties

The elastic constants are the physical parameters for testing the mechanical stability and strength of a solid and to determine the nature of a material's interatomic bondings. Information regarding elastic constants is also essential to assess the potential of a solid for mechanical and engineering applications. Since the compounds CaZn$_2$X$_2$ have trigonal crystal structures we obtained six independent elastic constants ($C_{ij}$): $C_{11}$, $C_{33}$, $C_{44}$, $C_{12}$, $C_{13}$ and $C_{14}$. These elastic constants are used to determine the various other elastic properties of the compounds. The single crystal elastic constants of the chosen compounds are listed in Table 2. A trigonal crystal system needs to meet the Born-Huang [26] criteria in order to be mechanically stable. The criteria are:

$$C_{11} - C_{12} > 0;$$

$$(C_{11} - C_{12})C_{44} - 2C_{14}^2 > 0;$$

$$(C_{11} + C_{12})C_{33} - 2C_{13}^2 > 0 \tag{2}$$

The calculated elastic constants satisfy all the above stability criteria at zero pressure. This indicates that the compounds $CaZn_2X_2$ (X = N, P, As) are mechanically stable. It is worth noticing that a small negative value of $C_{14}$ has no effect on the mechanical stability of the compound $CaZn_2N_2$. This actually suggests that this compound has a small internal strain.

**Table 2.** Single crystal elastic constants, $C_{ij}$ (in GPa), of $CaZn_2X_2$ (X = N, P, As) compounds.

| Compounds | $C_{11}$ | $C_{33}$ | $C_{44}$ | $C_{12}$ | $C_{13}$ | $C_{14}$ | Ref. |
|---|---|---|---|---|---|---|---|
| $CaZn_2N_2$ | 233.24 | 197.49 | 81.21 | 55.22 | 66.25 | -15.92 | This work |
|  | 242.10 | 195.70 | 81.90 | 60.70 | 72.30 | -15.90 | [7] |
| $CaZn_2P_2$ | 148.27 | 117.16 | 38.78 | 36.19 | 32.90 | 5.83 | This work |
| $CaZn_2As_2$ | 123.42 | 96.13 | 27.56 | 31.28 | 29.45 | 6.57 | This work |

The $C_{ij}$ of values obtained for $CaZn_2N_2$ in this study are in excellent agreement with the values estimated earlier [7]. The elastic constants for the other two compounds are new results. Among the six elastic constants, $C_{11}$ and $C_{33}$ describe the response to the uniaxial strain while the elastic constant $C_{44}$ connected to the shape deforming effects. $C_{12}$, $C_{13}$ and $C_{14}$ are off-diagonal elastic constants also linked with shape deforming shearing strain produced by stresses on different crystal planes. It is noticed from the Table 2 that the unidirectional elastic constants $C_{11}$ and $C_{33}$ are much higher than the pure shear elastic constant $C_{44}$. These results indicate that the shear deformation of the compounds is much easier than the unidirectional strain along any of the three crystallographic directions in the sequence of $CaZn_2N_2 > CaZn_2P_2 > CaZn_2As_2$. It is also noticed from the Table 2 that the value of $C_{11}$ is greater than $C_{33}$ for all the three compounds. Thus, it can be said that these three compounds are more compressible in the *c*-direction than in the *a(b)*-direction. It also suggests that the chemical bonding between nearest neighbors is stronger in the (100) planes than in the (001) planes. The large difference between the values of $C_{11}$ and $C_{33}$ also implies that the compounds' elastic characteristics are significantly anisotropic. The difference ($C_{12}$ - $C_{44}$) is referred to as the Cauchy pressure, and it is widely used to evaluate the nature of the chemical bonding in crystalline solids [27]. From Table 2 it is seen that the compounds $CaZn_2N_2$ and $CaZn_2P_2$ have a negative Cauchy pressure while the material $CaZn_2As_2$ has a positive Cauchy pressure. The negative value of Cauchy pressure implies that the compounds $CaZn_2N_2$ and $CaZn_2P_2$ have directional covalent bonding with angular character, whereas the positive Cauchy pressure of the compound $CaZn_2As_2$ suggests that its chemical bonding is dominated by ionic bonding which is spherically symmetric.

By using the independent elastic constants we have determined the different types of elastic moduli, Pugh's ratio, and Poisson's ratio of the compounds. The polycrystalline values of bulk modulus ($B$), shear modulus ($G$), Young's modulus ($Y$) and Poisson's ratio ($\eta$) are calculated by using the following equations:

$$B = \frac{B_V + B_R}{2} \qquad (3)$$

$$G = \frac{G_V + G_R}{2} \qquad (4)$$

$$Y = \frac{9BG}{3B + G} \qquad (5)$$

$$\eta = \frac{3B - 2G}{2(3B + G)} \qquad (6)$$

In these equations, $B_V$, $B_R$ and $G_V$, $G_R$ are the Voigt [28] and Reuss [29] values of bulk and shear modulus, respectively. The measured values of polycrystalline bulk modulus ($B$), shear modulus ($G$), Pugh's ratio ($B/G$), Young's modulus ($Y$) and Poisson's ratio ($\eta$) are given in Table 3.

The bulk and shear modulus can be used to estimate the mechanical properties of crystals. The bulk modulus of a material describes its ability to resist volume deformation. The shear modulus, on the other hand, predicts the material's capability to resist shape change. It is seen from the Table 3 that for all the compounds, the values of $B$ are higher than $G$. This higher value of $B$ suggests that shearing strain controls the mechanical stabilities of the $CaZn_2X_2$ compounds. The ductile or brittle nature of solids can be understood by using the values of Pugh's ratio ($B/G$). Pugh predicted that if the $B/G$ ratio is larger than 1.75, the material will show ductile behavior; otherwise, the material will exhibit brittleness [30]. Since all our compounds have $B/G$ ratio less than 1.75, it can be estimated that they will show brittleness.

**Table 3.** Elastic moduli (all in GPa), Pugh's ratio, and Poisson's ratio of the compounds $CaZn_2X_2$ (X = N, P, As).

| Compound | $B$ | $G$ | $B/G$ | $Y$ | $\eta$ | Ref. |
|---|---|---|---|---|---|---|
| CaZn$_2$N$_2$ | 115.33 | 80.59 | 1.43 | 196.09 | 0.22 | This work |
|  | 120.80 | 81.00 | 1.49 | 198.50 | 0.23 | [7] |
| CaZn$_2$P$_2$ | 68.17 | 46.52 | 1.47 | 113.70 | 0.22 | This work |
| CaZn$_2$As$_2$ | 57.74 | 35.60 | 1.62 | 88.60 | 0.24 | This work |

The stiffness and thermal shock resistance of the materials can be predicted using the calculated values of Young's modulus (*Y*). It is observed from the Table 3 that the value of *Y* is decreasing as we change the anion X from N to As. The low value of *Y* indicates that the compound CaZn$_2$As$_2$ is not stiff compared to the compounds CaZn$_2$N$_2$ and CaZn$_2$P$_2$. Therefore, the softness order of the materials can be written from the Table 3 as CaZn$_2$As$_2$ > CaZn$_2$P$_2$ > CaZn$_2$N$_2$. Again, it is known that the critical thermal shock resistance changes inversely with the Young's modulus [31]. Thus, the material CaZn$_2$As$_2$ is expected to show better thermal shock resistance than other two materials. All the elastic constants and moduli show an anti-correlation with the cell volume. A larger cell volume compound in a series has a smaller elastic strength.

The Poisson's ratio can be used to determine the nature of bonding forces and to evaluate the different mechanical properties of crystals. It can also be used to explore the stability of solids against shear [32]. If a crystalline solid has a low Poisson's ratio, it has structural stability against shear. All the three compounds in this work have low Poisson's ratio. Thus, it can be concluded that our materials have structural stability against shear. For central force solids, the lower and upper limits of the Poisson's ratio are 0.25 and 0.50 [33]. Table 3 shows that the Poisson's ratios of our compounds are outside of this range. Therefore, it can be said that these crystal systems are stabilized by non-central type forces. Moreover, the ductile or brittle nature of a crystal can also be predicted by the Poisson's ratio. A crystal with Poisson's ratio $\eta > 0.26$ is said to be ductile, whereas one with $\eta < 0.26$ is said to be brittle [31]. This criterion indicates that the compounds CaZn$_2$X$_2$ should exhibit brittleness. This result is consistent with the result predicted from the Pugh's ratio. Moreover, Poisson's ratio also represents a solid's plasticity against shear. The greater the Poisson's ratio, the greater is the plasticity. Hence, the compound CaZn$_2$As$_2$ has much better plasticity than the other two compounds.

The machinability index ($\mu^M$), the Kleinman parameter ($\zeta$), the bulk modulus along *a-*, *b-* and *c*-axis (known as directional bulk modulus) and isotropic bulk modulus ($B_{relax}$) are calculated using the following widely used equations [34,35] and the obtained values are enlisted in Table 4.

$$\mu^M = \frac{B}{C_{44}} \tag{7}$$

$$\zeta = \frac{C_{11} + 8C_{12}}{7C_{11} + 2C_{12}} \tag{8}$$

$$B_a = a\frac{dP}{da} = \frac{\Lambda}{1+\alpha+\beta} \tag{9}$$

$$B_b = b\frac{dP}{db} = \frac{B_a}{\alpha} \tag{10}$$

$$B_c = c\frac{dP}{dc} = \frac{B_a}{\beta} \quad (11)$$

$$B_{relax} = \frac{\Lambda}{(1+\alpha+\beta)^2} \quad (12)$$

with, $\Lambda = C_{11}+2C_{12}\alpha+C_{22}\alpha^2+2C_{13}\beta+C_{33}\beta^2+2C_{23}\alpha\beta$

$$\alpha = \frac{\{(C_{11}-C_{12})(C_{33}-C_{13})\}-\{(C_{23}-C_{13})(C_{11}-C_{13})\}}{\{(C_{33}-C_{13})(C_{22}-C_{12})\}-\{(C_{13}-C_{23})(C_{12}-C_{23})\}}$$

$$\beta = \frac{\{(C_{22}-C_{12})(C_{11}-C_{13})\}-\{(C_{11}-C_{12})(C_{23}-C_{12})\}}{\{(C_{22}-C_{12})(C_{33}-C_{13})\}-\{(C_{12}-C_{23})(C_{13}-C_{23})\}}$$

**Table 4.** The machinability index ($\mu^M$), Kleinman parameter ($\zeta$), bulk modulus ($B_{relax}$ in GPa), bulk modulus along $a$-, $b$- and $c$-axis ($B_a$, $B_b$ and $B_c$ in GPa), and $\alpha$ and $\beta$ of the CaZn$_2$X$_2$ (X = N, P, As) compounds.

| Compound | $\mu^M$ | $\zeta$ | $B_{relax}$ | $B_a$ | $B_b$ | $B_c$ | $\alpha$ | $\beta$ |
|---|---|---|---|---|---|---|---|---|
| CaZn$_2$N$_2$ | 1.42 | 0.39 | 145.86 | 465.29 | 465.29 | 391.00 | 1.0 | 1.19 |
| CaZn$_2$P$_2$ | 1.76 | 0.39 | 88.15 | 300.59 | 300.59 | 213.18 | 1.0 | 1.41 |
| CaZn$_2$As$_2$ | 2.10 | 0.40 | 73.55 | 253.03 | 253.03 | 175.72 | 1.0 | 1.44 |

A material's dry lubricating property can be evaluated from the machinability index ($\mu^M$). A material with high value of $\mu^M$ should show excellent lubricating properties, i.e., lesser friction. More importantly, machinability index measures the ease with which a solid can be cut and put into desired shapes. This parameter is important in materials engineering. From Table 4 it is evident that the compound CaZn$_2$As$_2$ should exhibit good machinability and better dry lubricity than the other two. It is suggestive to note that the machinability indices of CaZn$_2$X$_2$ (X = N, P, As) compounds are comparable to many technologically prominent MAX and MAB phase materials [36-40]. The Kleinman parameter ($\zeta$) is used to check the stability of a compound against stretching and bending type of strains [41]. This parameter has no dimension. The lower and upper limit of $\zeta$ is 0 and 1. The lower limit of $\zeta$ suggests a large contribution of bond stretching or contracting to resist external stress, while the upper limit denotes a significant contribution of bond bending to resist external load. Therefore, from Table 4 it can be inferred that mechanical strength in CaZn$_2$X$_2$ is primarily generated from the bond stretching/contracting contribution.

Again, from Table 4 it is observed that the value of directional bulk modulus $B_c$ is small compared to $B_a$ and $B_b$. This implies that $CaZn_2X_2$ is more compressible when stressed along the *c*-direction rather than along the *a*- or *b*-direction.

Information on the hardness of a solid is needed for the understanding its behavior in practical heavy duty applications. There are several theoretical schemes for calculating the hardness as a function of bulk modulus (*B*), shear modulus (*G*), Young's modulus, and Poisson's ratio. The formulae given by X. Chen et al. [42], Y. Tian et al. [43], and D. M. Teter [44] are based on either *G* or both *G* and *B*, but the formulae presented by N. Miao et al. [45] and E. Mazhnik et al. [46] are primarily dependent on the Young's modulus and Poisson's ratio. These formulas are shown below and the calculated values of hardness are enlisted in Table 5:

$$(H_V)_{Chen} = 2[\left(\frac{G}{B}\right)^2 G]^{0.585} - 3 \tag{13}$$

$$(H_V)_{Tian} = 0.92(G/B)^{1.137} G^{0.708} \tag{14}$$

$$(H_V)_{Teter} = 0.151 G \tag{15}$$

$$(H_V)_{Miao} = \frac{(1-2\eta)Y}{6(1+\eta)} \tag{16}$$

$$(H_V)_{Mazhnik} = \gamma_0 \chi(\eta) Y \tag{17}$$

where, $\chi(\eta)$ is a function of the Poisson's ratio and can be written as:

$$\chi(\eta) = \frac{1 - 8.5\eta + 19.5\eta^2}{1 - 7.5\eta + 12.2\eta^2 + 19.6\eta^3}$$

and $\gamma_0$ is a dimensionless constant with a value of 0.096.

**Table 5.** Calculated hardness (in GPa) with different schemes based on elastic moduli and Poisson's ratio for the $CaZn_2X_2$ (X = N, P, As) compounds.

| Compound | $(H_V)_{Chen}$ | $(H_V)_{Tian}$ | $(H_V)_{Teter}$ | $(H_V)_{Miao}$ | $(H_V)_{Mazhnik}$ |
|---|---|---|---|---|---|
| $CaZn_2N_2$ | 14.14 | 13.69 | 12.17 | 15.00 | 9.22 |
| $CaZn_2P_2$ | 9.09 | 9.03 | 7.02 | 8.70 | 5.35 |
| $CaZn_2As_2$ | 6.18 | 6.66 | 5.38 | 6.19 | 4.08 |

The formalism proposed by Mazhnik et al. [46] results in the lowest values of hardness for all three compounds. The qualitative trend is same in all the different approaches. It is seen from Table 5 that the hardness values obtained by different method are in the following order: $CaZn_2N_2 > CaZn_2P_2 > CaZn_2As_2$. Therefore, the compound $CaZn_2As_2$ is significantly softer than the compounds $CaZn_2N_2$ and $CaZn_2As_2$. $CaZn_2N_2$ is a fairly hard solid.

Anisotropy in elasticity influences a wide range of physical phenomena, including the production of microcracks in materials, crack motion, the development of plastic deformations in crystals, and so on. If a material's properties vary in different directions, it is said to be an anisotropic material. Understanding elastic anisotropy has important consequences in both crystal physics and applied engineering sciences. As a result, it is vital to estimate the elastic anisotropy parameters of $CaZn_2X_2$ in detail in order to gain information on its durability and potential uses under various external stress conditions. In this work, we determined the Zener anisotropy factor ($A$), shear anisotropy factors ($A_1$, $A_2$, $A_3$), and percentage anisotropy in compressibility ($A_B$) and shear ($A_G$). Calculated values are disclosed in Table 6. Futhermore, we have also calculate the universal anisotropy factor ($A^U$, $d_E$), equivalent Zener anisotropy factor ($A^{eq}$), the universal log-Euclidean index ($A^L$), the anisotropies of the bulk modulus along $a$- and $c$-axis ($A_{Ba}$ and $A_{Bc}$) and those are listed in Table 7. The following equations are used to calculate these anisotropy factors [32,34]:

$$A = \frac{2C_{44}}{C_{11}-C_{12}} \tag{18}$$

$$A_1 = \frac{4C_{44}}{C_{11}+C_{33}-2C_{13}} \tag{19}$$

$$A_2 = \frac{4C_{55}}{C_{22}+C_{33}-2C_{23}} \tag{20}$$

$$A_3 = \frac{4C_{66}}{C_{11}+C_{22}-2C_{12}} \tag{21}$$

$$A_B = \frac{B_V - B_R}{B_V + B_R} \tag{22}$$

$$A_G = \frac{G_V - G_R}{G_V + G_R} \tag{23}$$

$$A^U = \frac{B_V}{B_R} + 5\frac{G_V}{G_R} - 6 \geq 0 \tag{24}$$

$$d_E = \sqrt{A^U + 6} \tag{25}$$

$$A^{eq} = \left(1 + \frac{5}{12}A^U\right) + \sqrt{\left(1 + \frac{5}{12}A^U\right)^2 - 1} \tag{26}$$

$$A^L = \sqrt{\left[\ln\left(\frac{B_V}{B_R}\right)\right]^2 + 5\left[\ln\left(\frac{C_{44}^V}{C_{44}^R}\right)\right]^2} \tag{27}$$

with,

$$C_{44}^R = \left(\frac{5}{3}\right)\frac{C_{44}(C_{11} - C_{12})}{3(C_{11} - C_{12}) + 4C_{44}} \quad \text{is the Reuss value of } C_{44}$$

$$C_{44}^V = C_{44}^R + \left(\frac{3}{5}\right)\frac{(C_{11} - C_{12} - 2C_{44})^2}{3(C_{11} - C_{12}) + 4C_{44}} \quad \text{is the Voigt value of } C_{44}$$

$$A_{Ba} = \frac{B_a}{B_b} = \alpha \tag{28}$$

$$A_{Bc} = \frac{B_c}{B_b} = \frac{\alpha}{\beta} \tag{29}$$

**Table 6.** The Zener anisotropy factor ($A$), shear anisotropy factors ($A_1$, $A_2$, $A_3$), and percentage anisotropy in compressibility ($A_B$) and shear ($A_G$) of the CaZn$_2$X$_2$ (X = N, P, As) compounds.

| Compound | $A$ | $A_1$ | $A_2$ | $A_3$ | $A_B$ | $A_G$ |
|---|---|---|---|---|---|---|
| CaZn$_2$N$_2$ | 0.91 | 1.09 | 1.09 | 1 | 0.001 | 0.018 |
| CaZn$_2$P$_2$ | 0.70 | 0.78 | 0.78 | 1 | 0.007 | 0.021 |
| CaZn$_2$As$_2$ | 0.60 | 0.69 | 0.69 | 1 | 0.007 | 0.042 |

**Table 7.** The universal anisotropy factor ($A^U$, $d_E$), equivalent Zener anisotropy factor ($A^{eq}$), universal log-Euclidean index ($A^L$), anisotropies of the bulk modulus along $a$- and $c$-axis ($A_{Ba}$ and $A_{Bc}$) of the CaZn$_2$X$_2$ (X = N, P, As) compounds.

| Compound | $A^U$ | $d_E$ | $A^{eq}$ | $A^L$ | $A_{Ba}$ | $A_{Bc}$ |
|---|---|---|---|---|---|---|
| CaZn$_2$N$_2$ | 0.19 | 2.49 | 1.48 | 0.014 | 1 | 0.84 |
| CaZn$_2$P$_2$ | 0.23 | 2.50 | 1.54 | 0.210 | 1 | 0.71 |
| CaZn$_2$As$_2$ | 0.45 | 2.54 | 1.83 | 0.400 | 1 | 0.69 |

If the Zener anisotropy factor ($A$) and the shear anisotropy factors ($A_1$, $A_2$, $A_3$) of a crystal are equal to one then it is an isotropic crystal and if the factors are greater or lower than unity then the crystal will possess a degree of anisotropy. The calculated values of $A$, $A_1$ and $A_2$ in Table 6 are clearly different from unity, indicating that the compounds CaZn$_2$X$_2$ are anisotropic with regard to shearing stress along various crystal planes. On the other hand, it is seen that the shear anisotropy factor $A_3 = 1$ for all the studied compounds, which implies that the compounds are isotropic in nature for {001} shear planes between [110] and [010] directions. From the calculated percentage values of anisotropy in compressibility ($A_B$) and shear ($A_G$), it can be estimated that the compound $CaZn_2N_2$ is less anisotropic than the other two compounds in both compression and shear. It is known that the universal anisotropy factor, $A^U = 0$ for elastically isotropic crystals, whereas any other positive value indicates anisotropy [47]. For all the compounds in this study the values of $A^U$ are found to be greater than zero which suggests that the compound will exhibit large anisotropy. Again, the equivalent Zener anisotropy factor ($A^{eq}$) is unity for an isotropic crystal while any other value shows anisotropy. Thus, from Table 7 it can be suggested that all the compounds under study are anisotropic in nature. The value of the universal log-Euclidean index ($A^L$) lies in the range $0 \leq (A^L) \leq 10.26$ for inorganic crystalline compounds and for perfect isotropic crystal, it must be equal to unity [48]. All the compounds in this work have the values of $A^L$ less than one. Moreover, the calculated values of $A_{Ba}$ and $A_{Bc}$ imply that the bulk moduli for $CaZn_2X_2$ compounds are anisotropic along the $c$-axis and isotropic along the $a$-axis.

As far as three-dimensional (3D) visualization of the elastic parameters is concerned, 3D direction dependent Young's modulus, shear modulus, linear compressibility, and Poisson's ratio should have spherical shapes for isotropic crystals and any deviation from this shape would exhibit anisotropy. We present herein the 3D plots of the directional dependences of Young's modulus, shear modulus, linear compressibility (inverse of the bulk modulus), and Poisson's ratio for the compounds $CaZn_2X_2$ (X = N, P, As). These plots are generated by the ELATE [49] and shown in Figures 2 to 5.

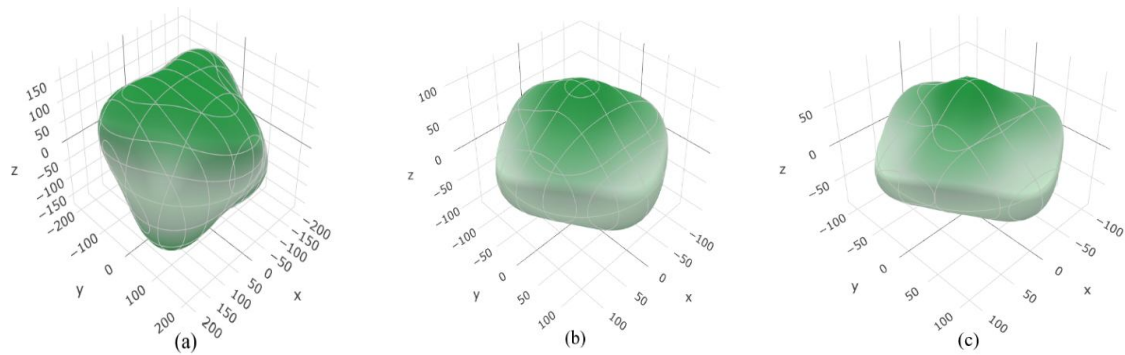

**Figure 2.** 3D directional dependences of Young's modulus for the compounds (a) $CaZn_2N_2$ (b) $CaZn_2P_2$ and (c) $CaZn_2As_2$.

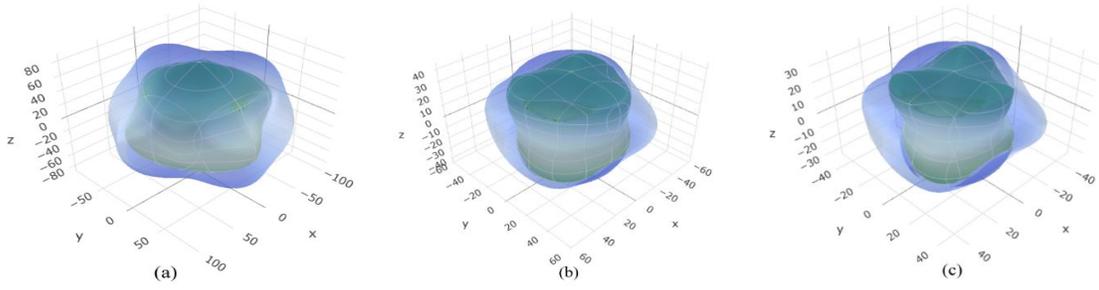

**Figure 3.** 3D directional dependences of shear modulus for the compounds (a) $CaZn_2N_2$ (b) $CaZn_2P_2$ and (c) $CaZn_2As_2$.

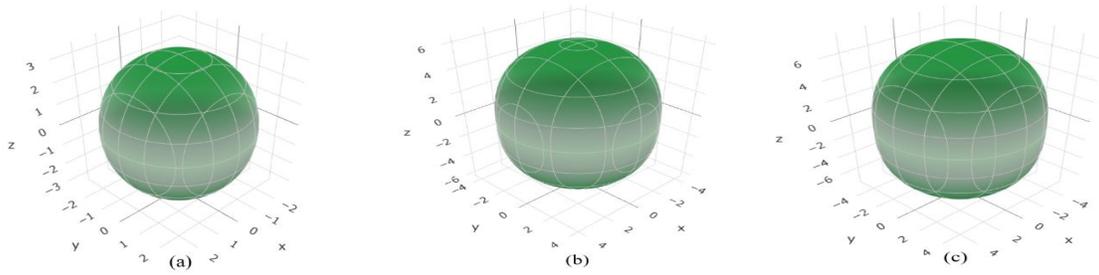

**Figure 4.** 3D directional dependences of linear compressibility for the compounds (a) $CaZn_2N_2$ (b) $CaZn_2P_2$ and (c) $CaZn_2As_2$.

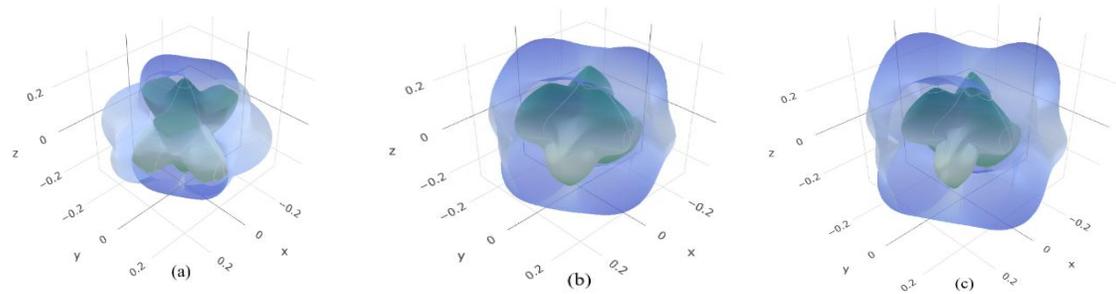

**Figure 5.** 3D directional dependences of Poisson's ratio for the compounds (a) $CaZn_2N_2$ (b) $CaZn_2P_2$ and (c) $CaZn_2As_2$.

Even though all the $CaZn_2X_2$ (X = N, P, As) compounds are isostructural, the degree of elastic anisotropy varies. Among the three solids, $CaZn_2As_2$ seems to possess the highest level of elastic anisotropy.

**3.3. Phonon dynamics**

The phonon dispersion spectra along Γ-M-K-Γ-A-L-H-A in the Brillouin zone (BZ) and the phonon density of states (DOS) of the $CaZn_2X_2$ (X = N, P, As) compounds are computed using the functional GGA(PBE). This calculation has been done by means of a finite displacement approach based on density functional perturbation theory (DFPT) [50]. The properties of phonons are crucial for understanding the crystalline materials. The structural stability, phase transition, and role of vibrations in thermal characteristics of a material can be described from the phonon dispersion spectra [51]. Superconductivity below the critical temperature also depends strongly on the phonon spectrum and also on the electron-phonon coupling constant. The calculated phonon dispersion curves and phonon DOS along high symmetry directions within the BZ of our titled compounds $CaZn_2X_2$ are shown in Figure 6. No negative frequency branch is observed in the dispersion curves. This implies that the $CaZn_2X_2$ crystals are dynamically stable at zero pressure. Figure 6 also shows that the number of phonon branches is fifteen, with three acoustic modes and twelve optical modes. The three acoustic modes occur in the low frequency values while the twelve optical modes occur at the high frequency values with some overlaps. In general, acoustic modes with low frequencies are generated by the vibration of a heavy atom, whereas optical modes with high frequencies are generated by the vibration of a light atom. $CaZn_2X_2$ (X = N, P, As) compounds do not have any phonon gap. The optical branches extend to much higher energies in $CaZn_2N_2$ in comparison to the other two compounds. This occurs since N atom has low mass and the overall stiffness of $CaZn_2N_2$ is much higher.

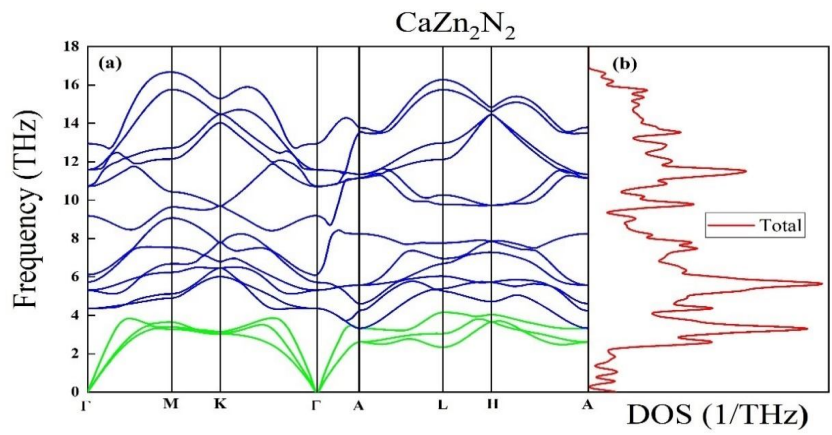

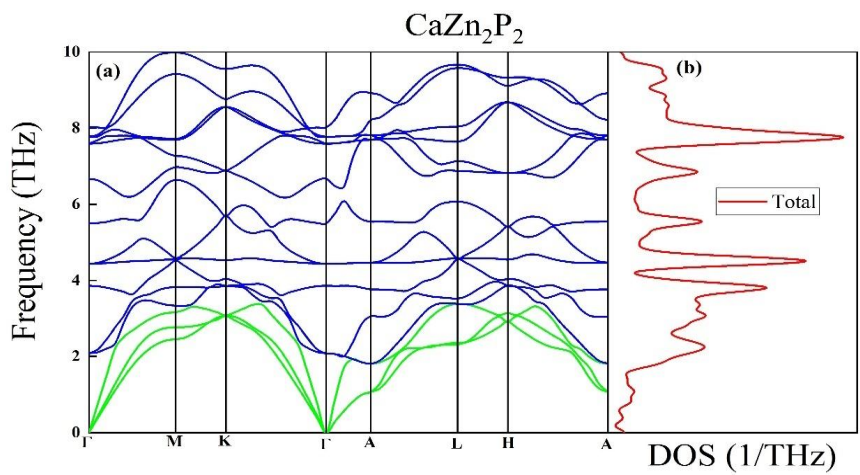

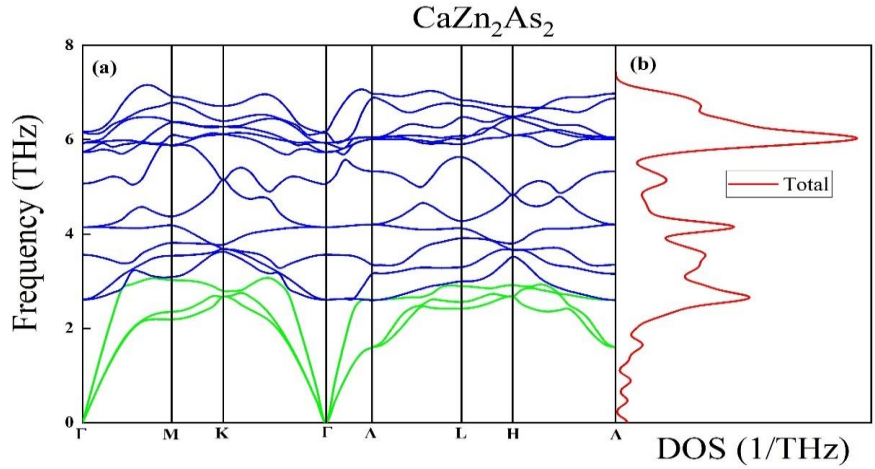

**Figure 6.** (a) Phonon dispersion spectra and (b) Total phonon DOS of the $CaZn_2X_2$ (X = N, P, As) compounds at zero pressure.

### 3.4. Electronic Properties

**(a) Band Structure**

In this study, we have used the HSE06 hybrid functional to calculate the band gaps of $CaZn_2X_2$ (X = N, P, As) compounds accurately. In certain cases the band gaps of the semiconductors are underestimated by the GGA and LDA type exchange-correlations functionals. In Figure 7, the calculated electronic band structures of $CaZn_2X_2$ compounds along several high symmetry directions in the first BZ are presented. From Figure 7 it is noticed that the compounds $CaZn_2N_2$ and $CaZn_2As_2$ exhibit direct band gaps at the $\Gamma$-point while the compound $CaZn_2P_2$ possesses an indirect band gap involving the $\Gamma$-M symmetry points. The energy band gap values of the $CaZn_2X_2$ compounds along with available experimental and theoretical data are presented in Table 8. It is seen from Table 8 that our computed energy band gaps match very well with the theoretical and experimental values. It is also seen from Table 8 that as the anion X is changed from N to As, the band gap values in $CaZn_2X_2$ compounds decrease. The decrease in band gap is due to the fact that the bond length increases while moving from N to As by substituting small size anion with large size anion.

**Table 8.** Calculated band gaps in eV of the $CaZn_2X_2$ (X = N, P, As) compounds with available theoretical and experimental values.

| Compound | Type of Band gap ($E_g$) | Theoretical $E_g$ [Ref.] | Experimental $E_g$ [Ref.] |
|---|---|---|---|
| $CaZn_2N_2$ | Direct | 1.79 [This work] | 1.90 [4] |
| | | 1.79 [11] | |
| $CaZn_2P_2$ | Indirect | 1.62 [This work] | - |
| | | 1.42 [11] | |
| $CaZn_2As_2$ | Direct | 1.10 [This work] | - |
| | Indirect | 1.27 [11] | |

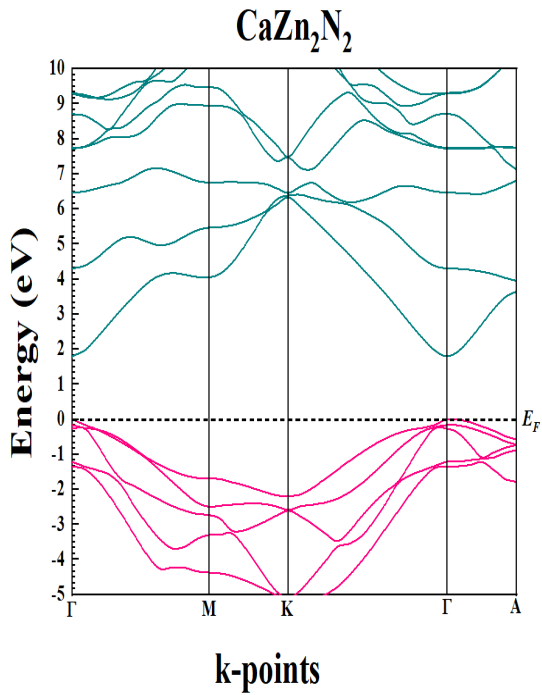
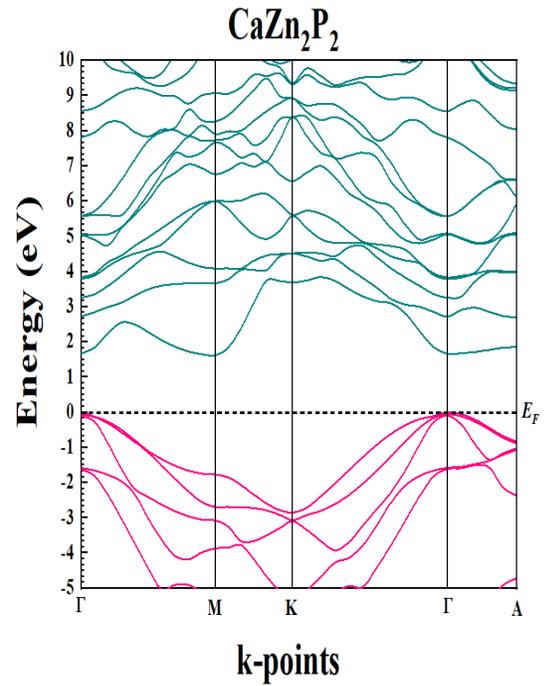

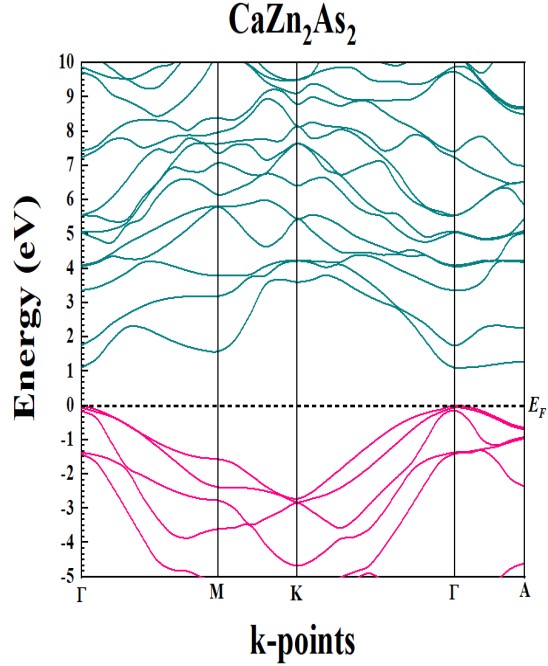

**Figure 7.** The calculated electronic band structures (using HSE06) of the $CaZn_2X_2$ (X = N, P, As) compounds along the high symmetry directions of the *k*-space within the first Brillouin zone.

Furthermore, it is noticed from Figure 7 that for all the three compounds, the band spectra along M-K and Γ-A directions show less dispersion while the spectra along Γ-M and K-Γ directions are highly dispersive. The less dispersive curves suggest that charge carriers have a large effective mass and, as a result, have little mobility in these directions. On the other hand, the high dispersive curves suggest low effective mass and high mobility of charge carriers.

**(b) Electronic energy density of states (DOS)**

To have a better hold of the electronic properties of the $CaZn_2X_2$ semiconductors, we have calculated the electronic energy density of states as a function of energy in the valence and conduction bands by using the energy dispersion curves. The computed total and partial density of states (TDOS and PDOS, respectively) of our chosen compounds are depicted in Figure 8. The position of a material's Fermi level and the magnitude of its TDOS at the Fermi energy are connected to its electronic and structural stability [52]. In Figure 8, the vertical broken line at 0 eV denotes the Fermi level, $E_F$. It is clearly seen from the figure that the TDOS of all the three compounds at Fermi level is zero, indicating that these compounds are semiconductors. It is also noticed from the Figure 8 that the energy separation between the valence band and the conduction band is significantly larger in $CaZn_2N_2$ than that in the other two compounds. Thus, this gap is reduced as we change the X anion from N to As. The PDOS plots illustrate the various contributions of electronic orbitals at different energy levels. It is observed from Figure 8 that near the Fermi level, the valence bands of $CaZn_2N_2$, $CaZn_2P_2$ and $CaZn_2As_2$ are mainly formed by the N-2$p$, P-3$p$ and As-4$p$ electronic states, respectively. Moreover, the conduction bands of these compounds above the Fermi level are mainly composed by the Zn-3$p$ and Zn-4$s$ electronic states. The bonding peak in TDOS is defined as the closest peak at negative energy below the Fermi level, whereas the anti-bonding peak is the closest peak at positive energy above the Fermi level [53,54]. It is seen from Figure 8 that the bonding and anti-bonding peaks are within 2 eV from the Fermi level for our compounds. This feature suggests that by using appropriate atomic replacement (alloying) or exerting pressure, it might be feasible to shift the electronic phase of the compounds under study. The closer proximity of the Fermi level to the bonding peak in CaZn$_2$N$_2$ contributes to its high overall crystal stiffness [55,56].

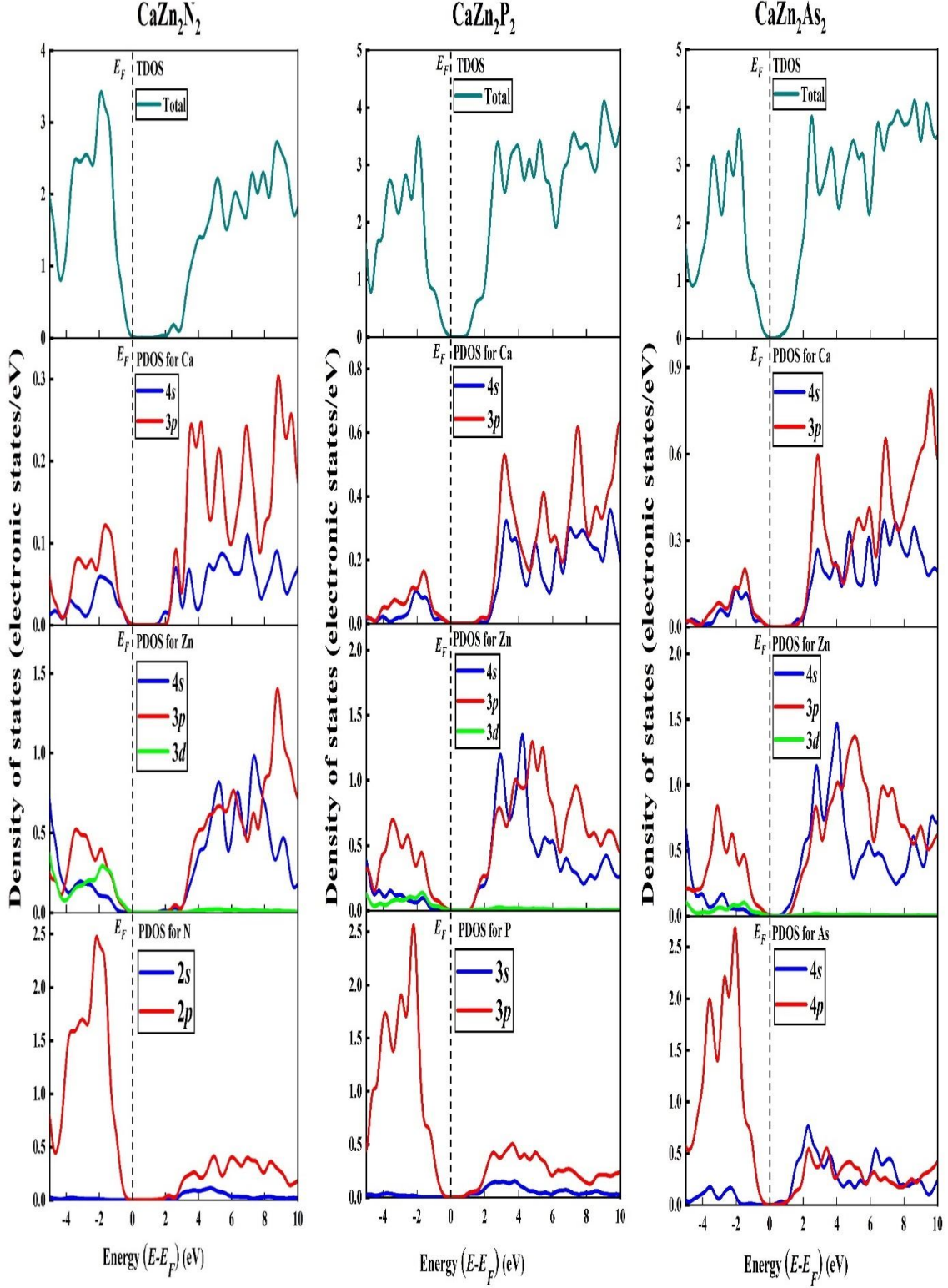

**Figure 8.** The computed total and partial density of states (TDOS and PDOS) plots of the $CaZn_2X_2$ (X = N, P, As) compounds.

### 3.5. Thermophysical properties

**(a) Debye temperature**

The thermal equivalent energy associated with the most energetic mode of vibration in a solid is known as the Debye temperature ($\theta_D$). This parameter is an essential quantity that is closely related to many thermophysical characteristics of solids including the thermal conductivity, lattice dynamics, melting temperature, coefficient of thermal expansion, elastic constants etc. In this work, we have calculated $\theta_D$ by applying the following equation [57]:

$$\theta_D = \frac{h}{k_B}\left[\left(\frac{3n}{4\pi V_0}\right)\right]^{\frac{1}{3}} v_m \qquad (30)$$

where, $h$ is the Planck's constant, $k_B$ is the Boltzmann's constant, $n$ denotes the number of atoms in the unit cell, $V_0$ is the equilibrium volume of the unit cell and $v_m$ is the average sound velocity in the solid. From the above equation, it is clear that $\theta_D$ is proportional to the mean sound velocity ($v_m$), which in turn depends on the elastic properties of a crystal. Now the value of $v_m$ can be estimated from bulk ($B$) and shear ($G$) modulus through longitudinal ($v_l$) and transverse ($v_t$) sound velocities by the following equations [58]:

$$v_m = \left[\frac{1}{3}\left(\frac{2}{v_t^3} + \frac{1}{v_l^3}\right)\right]^{-\frac{1}{3}} \qquad (31)$$

where,

$$v_t = \sqrt{\frac{G}{\rho}} \qquad (32)$$

$$v_l = \sqrt{\frac{3B+4G}{3\rho}} \qquad (33)$$

In (32) and (33), $\rho$ is the mass density. The calculated values of Debye temperatures for the compounds CaZn$_2$X$_2$, with $v_l$, $v_t$, $v_m$ and $\rho$ are listed in Table 9.

Table 9 shows that for all the three materials the values of longitudinal sound velocities are greater than the transverse sound velocities, and these velocities together with Debye temperatures decrease if one replaces the X anion from N to As. Moreover, the computed values of $v_l$, $v_t$ and $\theta_D$ for the compound CaZn$_2$N$_2$ are in excellent agreement with the previously estimated value [7].

**Table 9.** Calculated mass density ($\rho$ in kg m$^{-3}$), transverse, longitudinal and mean sound velocities ($v_t$, $v_l$ and $v_m$, all in m sec$^{-1}$), and Debye temperature ($\theta_D$ in K) of the CaZn$_2$X$_2$ (X = N, P, As) compounds.

| Compound | $\rho$ | $v_t$ | $v_l$ | $v_m$ | $\theta_D$ | Ref. |
|---|---|---|---|---|---|---|
| CaZn$_2$N$_2$ | 5043.08 | 3920.55 | 6518.50 | 4336.06 | 555.36 | This work |
|  | - | 3925.00 | 6599.00 | - | 556.40 | [7] |
| CaZn$_2$P$_2$ | 4019.28 | 3402.09 | 5691.49 | 3764.91 | 418.74 | This work |
| CaZn$_2$As$_2$ | 5025.12 | 2661.66 | 4575.60 | 2952.98 | 317.99 | This work |

**(b) Anisotropies in sound velocity**

Each atom in a solid has three vibrational modes: one longitudinal ($v_l$) and two transverse ($v_{t1}$ and $v_{t2}$). Sound propagates through a solid by exciting these vibrational modes. It is known that pure longitudinal and transverse modes in an anisotropic crystal are only supported along specific crystallographic directions while the modes of sound propagation in all other directions are quasi-transverse or quasi-longitudinal in nature. Since the three compounds under investigation have trigonal crystal structure with hexagonal symmetry, the acoustic velocities along the [100] and [001] directions can be calculated by using the equations [59] given below. The values thus obtained are disclosed in Table 10.

[100]:

$$[100]v_l = \sqrt{(C_{11} - C_{12})/2\rho} \; ; \; [010]v_{t1} = \sqrt{C_{11}/\rho} \; ; \; [001]v_{t2} = \sqrt{C_{44}/\rho} \qquad (34)$$

[001]:

$$[001]v_l = \sqrt{C_{33}/\rho}; \; [100]v_{t1} = [010]v_{t2} = \sqrt{C_{44}/\rho} \qquad (35)$$

here $v_{t1}$ and $v_{t2}$ are the first and second transverse modes, respectively.

**Table 10.** Anisotropic sound velocities (ms$^{-1}$) in the CaZn$_2$X$_2$ (X = N, P, As) compounds in different crystallographic directions.

| Propagation direction | | CaZn$_2$N$_2$ | CaZn$_2$P$_2$ | CaZn$_2$As$_2$ |
|---|---|---|---|---|
| [100] | $[100]v_l$ | 4120.27 | 3734.01 | 3027.86 |
|  | $[010]v_{t1}$ | 6669.73 | 6073.69 | 4955.87 |
|  | $[001]v_{t2}$ | 3935.60 | 3106.20 | 2341.89 |
| [001] | $[001]v_l$ | 6137.33 | 5399.03 | 4373.77 |
|  | $[100]v_{t1}$ | 3935.60 | 3106.20 | 2341.89 |
|  | $[010]v_{t2}$ | 3935.60 | 3106.20 | 2341.89 |

**(c) Lattice thermal conductivity**

Lattice vibrations are an effective avenue of heat transport. The ability of a material to transport heat his measured by the parameter known as the thermal conductivity. The lattice thermal conductivity ($k_{ph}$) of a solid at different temperatures determines the amount of heat energy carried by the lattice vibration when there is a temperature gradient in a solid. The $k_{ph}$ as a function of temperature can be calculated from a formula proposed by Slack [60] as follows:

$$k_{ph} = A \frac{M_{av} \Theta_D^3 \delta}{\gamma^2 n^{2/3} T} \quad (36)$$

here, $M_{av}$ is the average atomic mass in kg/mol, $\Theta_D$ is the Debye temperature in K, $\delta$ is the cubic root of the average atomic volume in meter, and $\gamma$ refers to the acoustic Grüneisen parameter, $n$ is to the number of atoms in the conventional unit cell and $T$ is the absolute temperature in K. The Grüneisen parameter measures the degree of anharmonicity of the phonons. The acoustic Grüneisen parameter and the factor $A$ ($\gamma$) due to Julian can be calculated by using the following relations [60,61]:

$$\gamma = \frac{3(1+\eta)}{2(2-3\eta)} \quad (37)$$

$$A(\gamma) = \frac{5.720 \times 10^7 \times 0.849}{2 \times (1 - 0.514/\gamma + 0.228/\gamma^2)} \quad (38)$$

here, $\eta$ refers to the Poisson's ratio. The lattice thermal conductivity is calculated at 300 K temperature for all the three crystals and the values obtained are tabulated in Table 11. It is clearly seen from the Table 11 that the values of $k_{ph}$ is decreasing as we move from N to As. This is expected since solids with higher Debye temperature have higher lattice thermal conductivities, as found in the preceding section. The overall thermal conductivities for all three compounds are low. Therefore, these materials can be used as efficient thermal barrier.

**(d) Minimum thermal conductivity and its anisotropy**

The behavior of a solid at temperatures higher than the Debye temperature has become a fascinating issue for high temperature applications. At high temperatures, a compound's inherent thermal conductivity approaches a lowest value which is called the minimum thermal conductivity ($k_{min}$). This parameter is important for the fact that it is not affected by the existence of defects in the crystal. Using the quasi-harmonic Debye model, Clarke developed the following equation for estimating the $k_{min}$ at high temperature [62]:

$$k_{min} = k_B v_m (V_{\text{atomic}})^{-\frac{2}{3}} \qquad (39)$$

In the above expression, $V_{atomic}$ is the cell volume per atom in the compound.

The values of minimum thermal conductivity calculated from Clarke formula for the semiconductors are given in Table 11. Again, we know that elastically anisotropic materials have different values of minimum thermal conductivity in different directions, i.e., they naturally exhibit anisotropy in minimum thermal conductivity. This anisotropy can be determined by the longitudinal and transverse sound velocities along different crystallographic directions. In this work, we used the following formula to compute the $k_{min}$ along various directions which is predicted by the Cahill model [63]:

$$k_{min} = \frac{k_B}{2.48} n^{2/3} (v_l + v_{t1} + v_{t2}) \qquad (40)$$

The calculated values of $k_{min}$ along [100] and [001] crystallographic directions are listed in Table 11.

**Table 11.** The lattice thermal conductivity $k_{ph}$ (W/m. K) at 300K, number of atoms per mole of the compound $n$ (m$^{-3}$), and minimum thermal conductivity $k_{min}$ (W/m. K) of the CaZn$_2$X$_2$ (X = N, P, As) compounds along different crystallographic directions.

| Compound | $k_{ph}$ | $n$ (10$^{28}$) | $k_{min}$ | | [100] $k_{min}$ | [001] $k_{min}$ |
|---|---|---|---|---|---|---|
| | | | Clarke | Cahill | | |
| CaZn$_2$N$_2$ | 31.55 | 7.94 | 1.11 | 1.48 | 1.51 | 1.44 |
| CaZn$_2$P$_2$ | 18.28 | 5.20 | 0.72 | 0.97 | 1.00 | 0.90 |
| CaZn$_2$As$_2$ | 9.95 | 4.72 | 0.53 | 0.71 | 0.75 | 0.66 |

From Table 11 it is seen that Clarke model predicts lower minimum thermal conductivity than Cahill model. It is also observed that the $k_{min}$ of CaZn$_2$N$_2$ compound is comparatively higher than the other two compounds. It can also be concluded from the above table that the $k_{min}$ along [100] direction is slightly higher than that of [001] direction. This implies small anisotropy in thermal transport.

**(e) Melting temperature**

A crystalline solid's thermal expansion, cohesive energy and the nature of atomic bonding are linked to its melting temperature. From the optimized elastic constants, one can calculate the melting temperature ($T_m$) of the $CaZn_2X_2$ (X = N, P, As) by using the following equation [64]:

$$T_m = 354 + 1.5(2C_{11} + C_{33}) \tag{41}$$

The obtained values of melting temperatures of our compounds are presented in Table 12. The highest melting temperature is found for $CaZn_2N_2$, while it is the lowest for the $CaZn_2As_2$ compound. These findings are in complete accord with the hardness, elastic moduli, and Debye temperature of $CaZn_2X_2$ (X = N, P, As) compounds.

**(f) Thermal expansion coefficient and heat capacity**

Thermal expansion coefficient (α) defines the extent in the dimensional change of a material due to a change in temperature. This particular parameter is associated with variety of physical properties, including thermal conductivity, specific heat, temperature dependence of the energy band gap for semiconductors, charge carrier effective mass etc. Thermal expansion coefficient can be computed from the following equation [65]:

$$\alpha = \frac{1.6 \times 10^{-3}}{G} \tag{42}$$

Thermal expansion coefficient is inversely related to the melting temperature of a solid as: $\alpha \approx 0.02 / T_m$ [62,66]. The calculated values of α for the compounds $CaZn_2X_2$ are shown in Table 12. The thermal expansion coefficient follows the sequence: $CaZn_2As_2 > CaZn_2P_2 > CaZn_2N_2$. The heat capacity is another useful thermophysical parameter. Materials with higher heat capacity offer higher thermal conductivity and lower thermal diffusivity. The change in thermal energy per unit volume of a material per degree Kelvin change in temperature is called the heat capacity per unit volume ($\rho C_P$). This parameter can be calculated from the following equation [65,66] for the compounds $CaZn_2X_2$ and the estimated values are presented in Table 12.

$$\rho C_P = \frac{3k_B}{\Omega} \tag{43}$$

In this equation, $\Omega$ is the volume per atom in the compound. The computed heat capacity per unit volume gives the value in the high temperature limit where the Dulong-Petit law holds.

The quantum of lattice vibrations is called a phonon. It affects a variety of physical characteristics, including electrical conductivity, thermal conductivity, and heat capacity. A compound's dominant phonon wavelength ($\lambda_{dom}$) is the wavelength at which the phonon distribution function becomes a maximum. In this work, we calculated the wavelength of the dominant phonon of our chosen materials at 300 K using the following expression [62,66]:

$$\lambda_{dom} = \frac{12.566 \, v_m}{T} \times 10^{-12} \tag{44}$$

The obtained values of $\lambda_{dom}$ for all the three compounds are tabulated in Table 12.

**Table 12.** The melting temperature $T_m$ (in K), thermal expansion coefficient α (in K$^{-1}$), heat capacity per unit volume $\rho C_P$ (in JK$^{-1}$m$^{-3}$) and wavelength of the dominant phonon mode at 300K $\lambda_{dom}$ (in m) of the CaZn$_2$X$_2$ (X = N, P, As) compounds.

| Compound | $T_m$ | α (10$^{-5}$) | $\rho C_P$ (10$^6$) | $\lambda_{dom}$ (10$^{-10}$) |
|---|---|---|---|---|
| CaZn$_2$N$_2$ | 1349.96 | 2.00 | 3.29 | 1.82 |
| CaZn$_2$P$_2$ | 974.55 | 3.44 | 2.15 | 1.58 |
| CaZn$_2$As$_2$ | 868.46 | 4.49 | 1.95 | 1.24 |

It is seen from Tables 5, 9 and 12 that among the compounds under study, the Debye temperature is maximum for the compound which has the highest melting temperature and the highest hardness. This is to be expected since a greater Debye temperature suggests stronger interatomic bonding, which results in a higher melting temperature and enhanced mechanical strength. It is also noticed that the wavelength of the dominant phonon is higher for the material CaZn$_2$N$_2$ since it has higher sound velocity and higher shear modulus than the other two compounds.

### 3.6. Bond population analysis

To learn more about the bonding characters of the CaZn$_2$X$_2$ (X = N, P, As) compounds, we have studied both Mulliken population analysis (MPA) [67] and Hirshfeld population analysis (HPA) [68]. The results of this analysis for the materials CaZn$_2$X$_2$ are presented in Table 13. Mulliken charge measures how the electronic structure changes with atomic displacement. This charge is also associated with dipole moment, polarizability, electronic structure, charge mobility in chemical reactions and other related properties of crystal systems. The charge spilling parameter represents the amount of missing valence charges in the projection. The smaller the value of the parameter, the better the modeling of electronic bonding. From the data given in Table 13 it is found that the charge spilling parameters of the CaZn$_2$X$_2$ (X = N, P, As) compounds are very low. This indicates a good representation of projections of atomic orbitals in the MPA. From this analysis, it is observed that the Zn atoms have a substantially higher total charge than the other atoms. This mainly comes from the 3$d$ states of the Zn atom. The effective valence charge (EVC) is defined as the difference between the formal ionic charge and the computed Mulliken charge [69]. For all the three compounds, the values of EVC are found to be non-zero which suggests covalent interaction between the atoms of the studied compounds. It is also noticed from Table 13 that the Mulliken charge and Hirshfeld charge is negative for the N, P and As atoms indicating that these atoms act as anions. Again, it is seen that the charge transfer between the Ca atom and the X (X = N, P, As) anions is much higher while the charge transfer between the Zn atom and X is very small. This feature suggests that the bonding among Ca-X is ionic while Zn-X possesses covalent bonding nature. There is qualitative agreement between the results obtained from the MPA and HPA. It can be concluded that the CaZn$_2$X$_2$ (X = N, P, As) compounds have mixed bonding features with both ionic and covalent contributions.

**Table 13.** Charge spilling parameter (%), orbital charges (electron), atomic Mulliken charge (electron), Hirshfeld charge (electron), and EVC (electron) of the $CaZn_2X_2$ (X = N, P, As) compounds.

| Compound | Atoms | Charge spilling (%) | s | p | d | Total | Mulliken charge | Hirshfeld charge | EVC |
|---|---|---|---|---|---|---|---|---|---|
| $CaZn_2N_2$ | Ca | 0.34 | 2.16 | 6.15 | 0.00 | 8.31 | 1.69 | 0.17 | 0.31 |
| | Zn | | 0.59 | 1.20 | 9.94 | 11.73 | 0.27 | 0.34 | 1.73 |
| | N | | 1.74 | 4.38 | 0.00 | 6.11 | -1.11 | -0.42 | -1.89 |
| $CaZn_2P_2$ | Ca | 0.38 | 2.22 | 6.40 | 0.00 | 8.62 | 1.38 | 0.13 | 0.62 |
| | Zn | | 0.52 | 1.37 | 9.96 | 11.84 | 0.16 | 0.24 | 1.84 |
| | P | | 1.66 | 4.19 | 0.00 | 5.85 | -0.85 | -0.30 | -2.15 |
| $CaZn_2As_2$ | Ca | 0.28 | 2.34 | 6.50 | 0.00 | 8.84 | 1.16 | 0.16 | 0.84 |
| | Zn | | 0.71 | 1.30 | 9.97 | 11.98 | 0.02 | 0.22 | 1.98 |
| | As | | 1.45 | 4.15 | 0.00 | 5.60 | -0.60 | -0.31 | -2.4 |

### 3.7. Charge density distribution

Mapping of electron charge density provides detailed information about a material's chemical bonding and charge transfer among its atoms. In this work, we have calculated the electronic charge density distribution of $CaZn_2X_2$ in (110) and (111) crystal planes. Figure 9 depicts the electronic charge density distribution. The color scale between the charge density maps represents the total electron density.

It is clearly seen from Figure 9 that weak charge sharing occurs between the Zn and X anions. Thus, covalent bonding dominates between these two atoms. On the other hand, there is no charge sharing between the Ca and X atoms which indicates that the bonding among Ca-X is of ionic nature. Therefore, the compounds $CaZn_2X_2$ have mixed covalent and ionic bonding nature, where the ionic contribution dominates. This result is consistent with the previously studied bond population analyses.

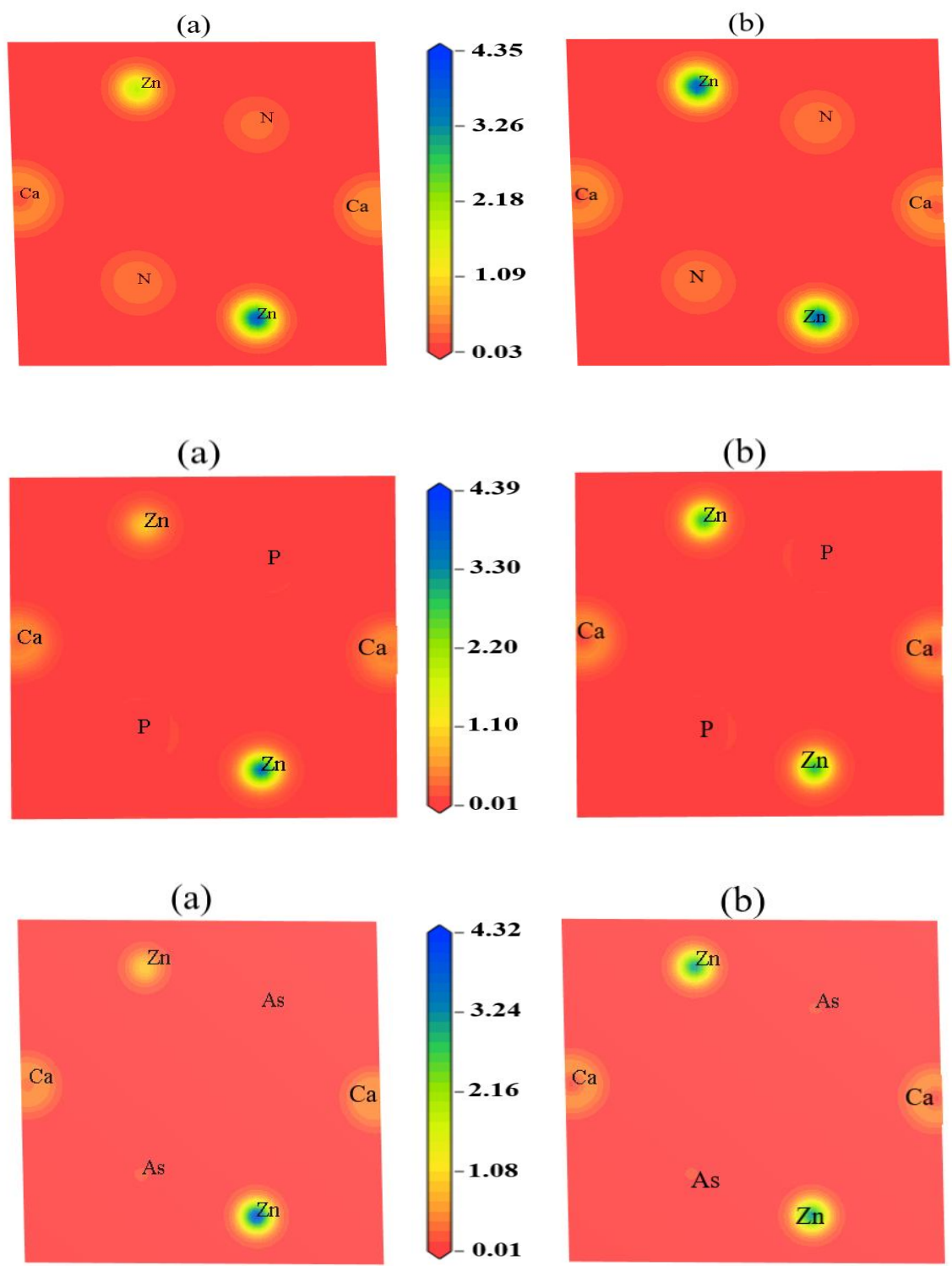

**Figure 9.** Charge density distribution map in the $CaZn_2X_2$ (X = N, P, As) crystals in the (a) (110) and (b) (111) planes. The charge density scale is given in between the panels.

### 3.8. Optical properties

A material's optical response can be determined by a number of energy/frequency dependent parameters, which include the dielectric function $\varepsilon(\omega)$, energy loss function $L(\omega)$, absorption coefficient $\alpha(\omega)$, refractive index $n(\omega)$, optical conductivity $\sigma(\omega)$ and reflectivity $R(\omega)$ (where $\omega$ is the angular frequency). These optical parameters are computed in this section and are presented in Figures 10, 11, and 12. Since the structural and elastic parameters of $CaZn_2X_2$ are direction dependent (i.e. anisotropic), optical parameters might also exhibit anisotropy. Therefore, we have computed the optical parameters of $CaZn_2X_2$ compounds for the incident electric fields in two different polarization directions, [100] and [001]. The optical spectra are shown for incident photon energies up to 25 eV.

The electronic nature of a material can be understood through the values of absorption coefficient $\alpha(\omega)$. The $\alpha(\omega)$ the $CaZn_2X_2$ (X = N, P, As) crystals are depicted in Figures. 10(a), 11(a), and 12(a). It is observed from these figures that the materials $CaZn_2N_2$, $CaZn_2P_2$, $CaZn_2As_2$ start absorption at approximately 1.5 eV, 1.2 eV and 0.9 eV energies, respectively. These values give the optical band gaps. This also confirms the fact that these compounds are semiconductors. Further observation of these figures reveals that as the anions change from N to As, the maximum peaks of $\alpha(\omega)$ decrease and shift to lower energies. It is interesting to note that all of the materials exhibit significant absorption in the ultraviolet region of the electromagnetic spectrum. The absorption coefficient is anisotropic with respect to the polarization states of the electric field.

The conductivity of free charge carriers over a specific range of photon energy is referred to as the optical conductivity. Figures 10(b), 11(b) and 12(b), reveal that the optical conductivity $\sigma(\omega)$ of $CaZn_2X_2$ compounds do not start from zero photon energy. It is consistent with the previously calculated electrical band structure and TDOS calculations and implies once again that the materials are semiconductors. Beyond the starting points for all the compounds, $\sigma(\omega)$ increases with photon energy and reaches their maximum value, then gradually decreases with further increase in energy. It is also noticed that the compound $CaZn_2N_2$ shows higher amount of anisotropy in optical conductivity than the other two compounds. Interestingly, when the anion is changed from N to As, the optical conductivity spectra become higher and move towards the lower energies.

Complex dielectric function is a vital parameter to describe a material's optical properties since the other energy dependent optical constants can be obtained from it. The variation of real (Re), $\varepsilon_1(\omega)$ and imaginary (Im), $\varepsilon_2(\omega)$ parts of dielectric function of $CaZn_2X_2$ is illustrated in Figures 10(c), 11(c) and 12(c). From these figures it is seen that the values of $\varepsilon_1(0)$ is increasing as we change the X anion from N to As. Therefore, it can be concluded that the static dielectric constant $\varepsilon_1(0)$ increases with the decrease of band gap. Above the zero frequency, $\varepsilon_1(\omega)$ starts to increase, reaches a maximum value, and then it abruptly decreases and reaches below zero. The imaginary part, which is related to the dielectric loss, has a close correspondence with the optical absorption coefficient. From the dielectric spectra, it is observed that the peak values of $\varepsilon_2(\omega)$ are increasing as we replace the X anion from N to As.

The absorption and reflection properties of a substance are related to its loss function. Additionally, the trailing edges of a material's reflection and absorption spectra correspond to the peak position in the loss function [70,71]. The energy loss function, $L(\omega)$, of the $CaZn_2X_2$ compounds is depicted in Figures 10(d), 11(d) and 12(d). These graphs showed that the sharp spectral peaks of $L(\omega)$ shifted towards the lower energy due to the switch from N to As anions. The positions of the loss peaks corresponds to the energy of the plasma excitations in the $CaZn_2X_2$ (X = N, P, As) compounds.

The reflectivity spectra, $R(\omega)$, of $CaZn_2X_2$ are displays in Figures 10(e), 11(e) and 12(e). The $R(\omega)$ spectra of the compounds increases in strength if we replace the anion X from N to As. Reflectivity is low and fairly anisotropic for the $CaZn_2N_2$ compound. The other two compounds have good reflectivity in the ultraviolet region and the optical anisotropy is also low.

The group velocity of an electromagnetic wave inside a material is determined by the real part of the refractive index, $n(\omega)$. On the other hand, the imaginary part (referred to as the extinction coefficient), $k(\omega)$ controls how much the incident electromagnetic wave is attenuated when it passes through the substance. The complex refractive index spectra of $CaZn_2X_2$ crystals are shown in Figures 10(f), 11(f) and 12(f). From these plots, it can be concluded that the $n(0)$ values are increasing as one goes from N to As. Thus, it can be said that the energy band gap, decreases from N to As, as it is inversely associated with the variation of $n(0)$. Moreover, it also noted that $n(0)$ increases significantly when the anion X is switched from N to As.

The overall results obtained in this section imply that the compound $CaZn_2N_2$ exhibits higher amount of optical anisotropy than the other two compounds. Furthermore, our calculated results are in good agreement with the previously calculated results by Murtaza et al. [11].

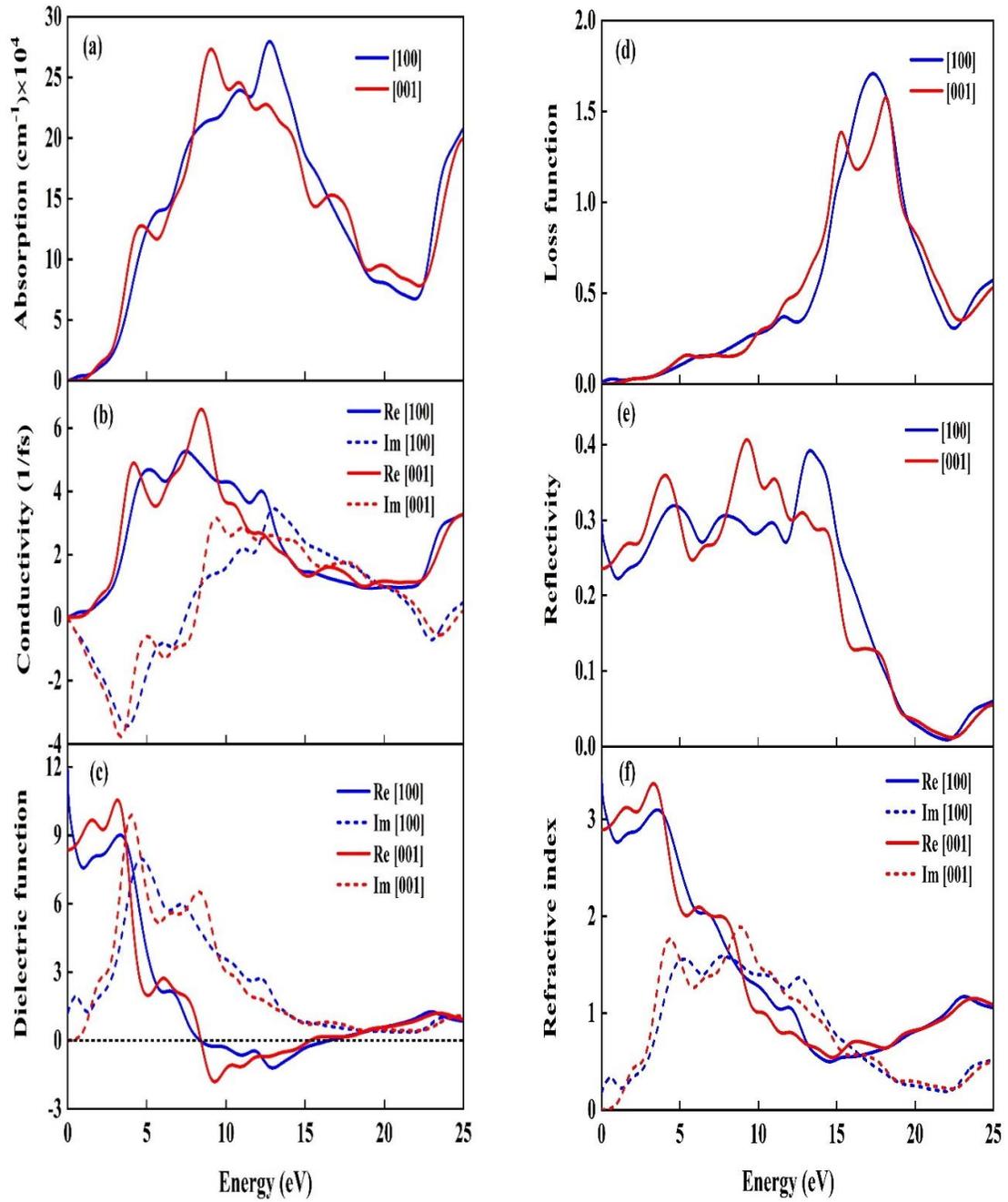

**Figure 10.** The energy dependent (a) absorption coefficient, (b) optical conductivity, (c) dielectric function, (d) energy loss function, (e) reflectivity, and (f) refractive index of $CaZn_2N_2$ with electric field polarization vectors along [100] and [001] directions.

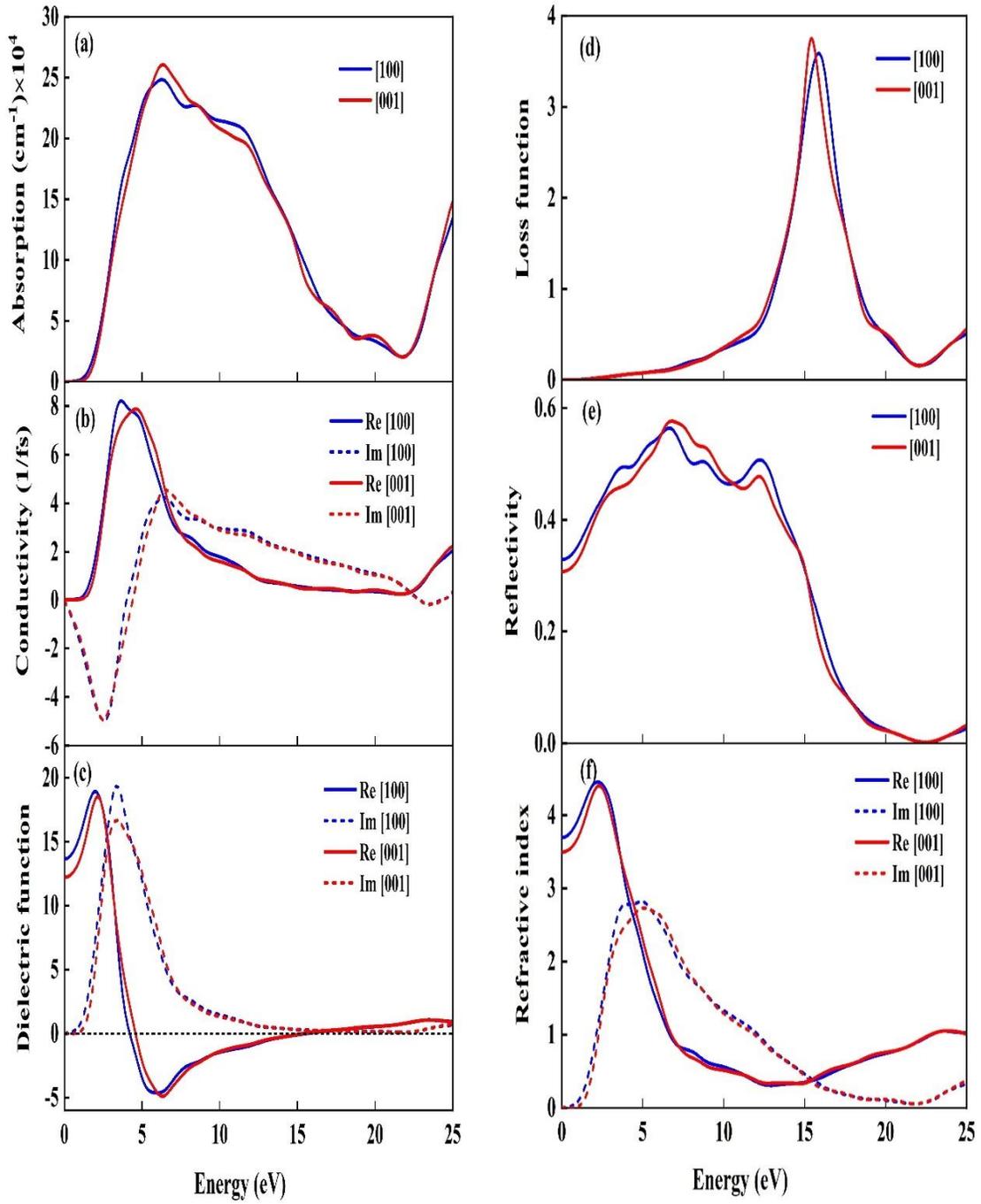

**Figure 11.** The energy dependent (a) absorption coefficient, (b) optical conductivity, (c) dielectric function, (d) energy loss function, (e) reflectivity, and (f) refractive index of $CaZn_2P_2$ with electric field polarization vectors along [100] and [001] directions.

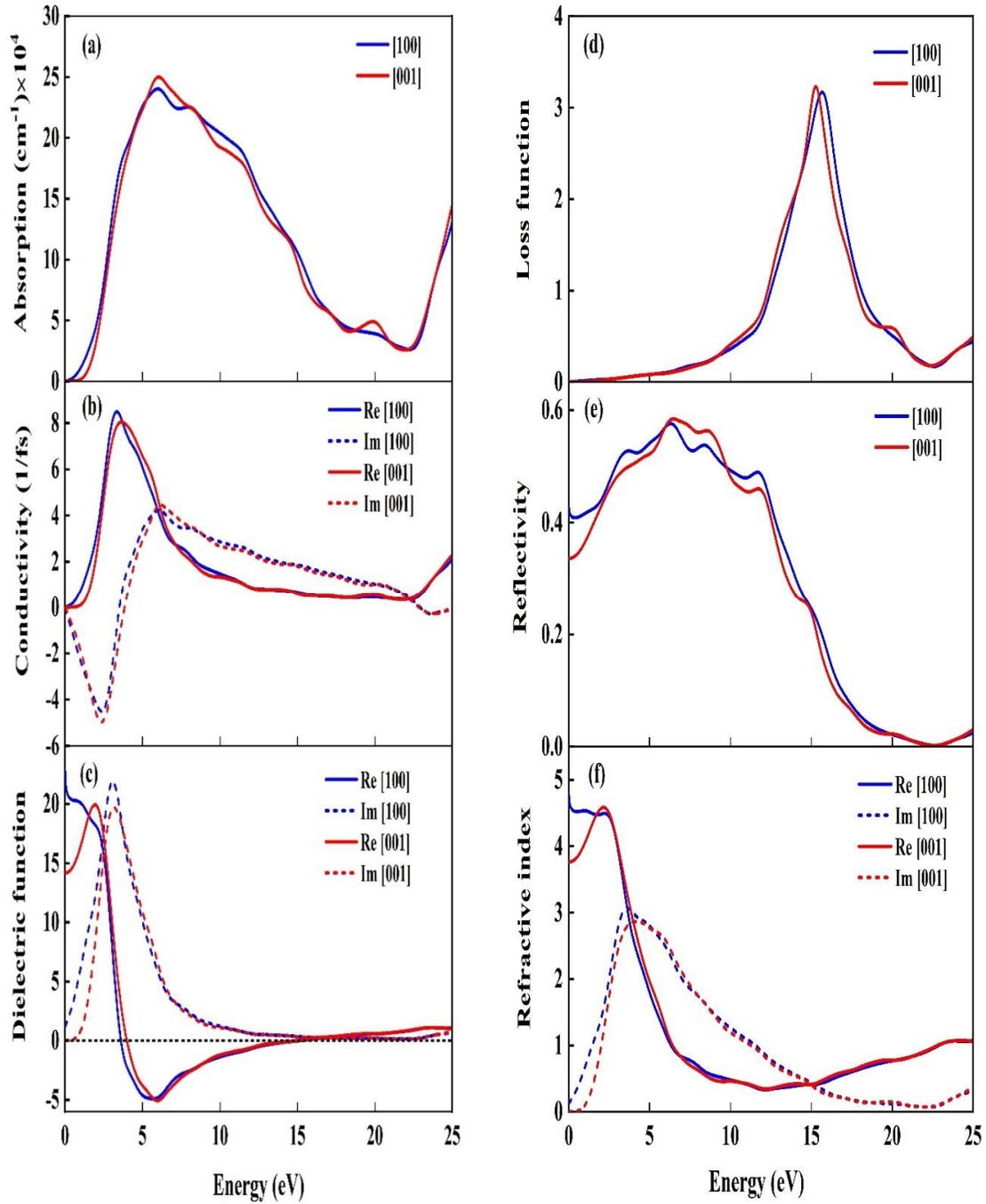

**Figure 12.** The energy dependent (a) absorption coefficient, (b) optical conductivity, (c) dielectric function, (d) energy loss function, (e) reflectivity, and (f) refractive index of $CaZn_2As_2$ with electric field polarization vectors along [100] and [001] directions.

## 4. Conclusions
The key findings of the current study are listed below.

- ❖ The structural optimization of our titled compounds has been accomplished at zero pressure and temperature. The optimized lattice parameters are in good agreement with those found in previous studies.
- ❖ When the anions X in $CaZn_2X_2$ compounds are altered from N to As, the lattice parameter of the compounds rises and the *c/a* ratio decreases. This is largely due to the variation in the ionic radii. Decreasing *c/a* ratio implies stronger interlayer bonding.
- ❖ All three compounds studied here are mechanically stable.
- ❖ Various elastic indicators show that the $CaZn_2X_2$ (X = N, P, As) crystals are brittle in nature. Among the three compounds investigated here, $CaZn_2N_2$ is the hardest and least machinable; $CaZn_2As_2$ is the softest and most machinable.
- ❖ The bulk moduli *(B)* of all three compounds are higher than the shear moduli *(G)*. Therefore, it can be anticipated that the shearing strain will govern the mechanical stabilities of the $CaZn_2X_2$ (X = N, P, As) compounds.
- ❖ According to the calculated values of different elastic anisotropy factors, we can conclude that these materials possess a degree of elastic anisotropy.
- ❖ From the phonon spectra, it is observed that all the materials under study exhibit positive phonon frequencies in the whole BZ. This result implies that the compounds under study are dynamically stable.
- ❖ Using the HSE06 scheme, we found that the compound $CaZn_2P_2$ has indirect band gap while the other two compounds $CaZn_2N_2$ and $CaZn_2As_2$ possess direct band gaps. Calculated band gap values reveal that these compounds hold promise to be used in photovoltaic applications.
- ❖ From the computed density of states (DOS) plots, it is seen that the bonding and anti-bonding peaks are within 2 eV from the Fermi level. This feature indicates that it would be possible to modify the electronic characters of our titled compounds by applying pressure or by alloying with the proper atomic species.
- ❖ The softness order found for these compounds is $CaZn_2As_2 > CaZn_2P_2 > CaZn_2N_2$, consistent with the calculated Debye temperature, melting point, and coefficient of thermal expansion.
- ❖ From the bond population analysis and charge density distribution it is found that the compounds have mixed covalent and ionic bonding nature.
- ❖ The optical parameters spectra correspond very well to the electronic band structure features. The compounds are efficient absorber and reflector of ultraviolet light. The low energy real part of the refractive index is high and increases in the sequence $CaZn_2As_2 > CaZn_2P_2 > CaZn_2N_2$. All these features are useful for optoelectronic device applications. The $CaZn_2X_2$ (X = N, P, As) compounds exhibit some optical anisotropy with $CaZn_2N_2$ showing higher level of optical anisotropy than the other two. We also observe that all the optical characteristics of the compounds change sequentially as the X anions move through N to As.

To summarize, a large number of physical properties of the $CaZn_2X_2$ (X = N, P, As) compounds have been explored theoretically in this paper. Many of the results presented are novel and should be treated as references for future studies. We hope that the results presented and discussed herein will inspire researchers to explore these compounds further in future.

**Acknowledgements**
S. H. N. and R. S. I. acknowledge the research grant (1151/5/52/RU/Science-07/19-20) from the Faculty of Science, University of Rajshahi, Bangladesh, that partly supported this work. S. H. N. dedicates this work to the loving memory of his father, Professor A. K. M. Mohiuddin, who passed away this year.

## Declaration of interest
The authors declare that they have no known competing financial interests or personal relationships that could have appeared to influence the work reported in this paper.

**Data availability**
The data sets generated and/or analyzed in this study are available from the corresponding author on reasonable request.

## CRediT author statement